\documentclass[11pt,journal]{IEEEtran}
\usepackage[utf8]{inputenc}
\usepackage{makecell,multirow,array}
\usepackage{hhline}
\usepackage{setspace}
\usepackage{threeparttable}
\usepackage{etoolbox}
\appto\TPTnoteSettings{\scriptsize}
\usepackage{float}
\usepackage{amssymb}
\usepackage{booktabs}
\usepackage{capt-of}
\usepackage{sidecap}
\usepackage{framed}

\newcolumntype{C}[1]{>{\centering\let\newline\\\arraybackslash\hspace{0pt}}m{#1}}
\usepackage{cite}

\usepackage[pdftex]{graphicx}
\graphicspath{ {./images/} }
\usepackage[export]{adjustbox}
\usepackage[caption=false,font=footnotesize,position=t,singlelinecheck=off,justification = raggedright]{subfig}

\usepackage{amsmath}
\DeclareMathOperator {\prediction}{prediction}
\DeclareMathOperator {\truelabel}{truelabel}

\DeclareMathOperator {\truevalue}{truevalue}
\DeclareMathOperator {\n}{n}
\usepackage[
  separate-uncertainty = true,
  multi-part-units = repeat
]{siunitx}
\usepackage{algorithm}
\usepackage{algpseudocode}

\usepackage{url}

\author{
Chengyang~Zhou\textsuperscript{1,2}, 
Thao~Vy~Dinh\textsuperscript{1,2}\IEEEauthorrefmark{1}, 
Heyi~Kong\textsuperscript{1,2}\IEEEauthorrefmark{1}, 
Jonathan~Yap\textsuperscript{3,4}, 
Khung~Keong~Yeo\textsuperscript{3,4},
Hwee~Kuan~Lee\textsuperscript{1,5,6,7,8}, 
and~Kaicheng~Liang\textsuperscript{1,9}

\vspace{\baselineskip}
\footnotesize{\textit{
\IEEEauthorblockA{\textsuperscript{1}Bioinformatics Institute, Agency for Science, Technology and Research (A*STAR), Singapore\\}
\IEEEauthorblockA{\textsuperscript{2}Hwa Chong Institution (College), Singapore\\}
\IEEEauthorblockA{\textsuperscript{3}National Heart Centre Singapore\\}
\IEEEauthorblockA{\textsuperscript{4}Duke-NUS Medical School, Singapore\\}
\IEEEauthorblockA{\textsuperscript{5}School of Computing, National University of Singapore\\}
\IEEEauthorblockA{\textsuperscript{6}Singapore Eye Research Institute (SERI)\\}
\IEEEauthorblockA{\textsuperscript{7}Image and Pervasive Access Laboratory (IPAL), Singapore\\}
\IEEEauthorblockA{\textsuperscript{8}Rehabilitation Research Institute of Singapore\\}
\IEEEauthorblockA{\textsuperscript{9}Institute of Bioengineering and Nanotechnology, A*STAR, Singapore\\}
\IEEEauthorblockA{\IEEEauthorrefmark{1}Equal contribution\\\vspace{-6ex}}
}}}
\title{Automated Deep Learning Analysis of Angiography \\ Video Sequences for Coronary Artery Disease\vspace{-1ex}}
\begin{document}
\maketitle
\begin{abstract}
The evaluation of obstructions (stenosis) in coronary arteries is currently done by a physician's visual assessment of coronary angiography video sequences. It is laborious, and can be susceptible to interobserver variation. Prior studies have attempted to automate this process, but few have demonstrated an integrated suite of algorithms for the end-to-end analysis of angiograms. We report an automated analysis pipeline based on deep learning to rapidly and objectively assess coronary angiograms, highlight coronary vessels of interest, and quantify potential stenosis. We propose a 3-stage automated analysis method consisting of key frame extraction, vessel segmentation, and stenosis measurement. We combined powerful deep learning approaches such as ResNet and U-Net with traditional image processing and geometrical analysis. We trained and tested our algorithms on the Left Anterior Oblique (LAO) view of the right coronary artery (RCA) using anonymized angiograms obtained from a tertiary cardiac institution, then tested the generalizability of our technique to the Right Anterior Oblique (RAO) view. We demonstrated an overall improvement on previous work, with key frame extraction top-5 precision of 98.4\%, vessel segmentation F1-Score of 0.891 and stenosis measurement 20.7\% Type I Error rate. 
\end{abstract}

\vspace{-3ex}
\section{Introduction}

\IEEEPARstart{C}{oronary} artery disease (CAD) was responsible for 18.1\% of deaths in Singapore in 2018 \cite{ministry}. Plaque buildup in the coronary arteries impedes blood flow and affects oxygen supply to the heart, which may cause chest pain and even heart attacks. Coronary angiography is a widely used and increasingly common interventional procedure used to assess the coronary arteries for potential plaque obstruction \cite{doi:10.1161/CIRCINTERVENTIONS.112.000205}. An iodinated radiopaque contrast agent is injected into the coronary arteries via a catheter, followed by a fluoroscopy scan that acquires a multi-frame video sequence. Due to the 2D nature of X-ray projections, scans are typically acquired from various angles to obtain more comprehensive visualization.

The interpretation of coronary angiograms requires specialized training and substantial experience in interventional cardiology, and can be time-consuming. As a limited amount of contrast agent dissipates in the vessels, only a limited portion of each fluoroscopy video offers image frames of sufficient quality (``key frames"). These are the frames that display the complete blood vessel of interest with high contrast, allowing cardiologists to make assessments on the degree of plaque obstruction (stenosis). The procedure can have substantial variation in key frame selection, as well as fluctuations in image contrast or visibility due to device angular placement or biological variations\cite{katritsis_limitations_1991}. Also, the current reliance on a physician's visual assessment of angiograms is susceptible to inter-observer variability; the same video interpreted as 50\% stenosis by one physician might be interpreted as 70\% by another \cite{MARCUS1988882}. A study of general cardiologists performing angiographic assessments reported a 95\% confidence interval of 22 percentage points \cite{kussmaul_accuracy_1992}, and even panels of experienced angiographers reading sequential angiograms taken 2 years apart reported substantial intrapanel and interpanel disagreement\cite{detre_reliability_1982}. These challenges in angiographic reading facing both general and experienced interventional cardiologists highlight the risk of missed/unnecessary interventional treatments, and the need for an objective and automated tool that enables rapid processing and analysis of angiography video sequences. Even resource-heavy core labs staffed with professional angiographers, which are often used in clinical research and trials evaluating patient outcomes, may suffer from the same challenges. 

Recent developments in image classification and segmentation with deep learning and neural networks have opened up new possibilities in medical image analysis. In this paper, we developed a deep learning-based set of algorithms for automated analysis of coronary angiography video sequences. Our 3-stage pipeline, which consisted of key frame extraction, vessel segmentation and stenosis measurement, simulated the clinical workflow and broke down the assessment process into explainable steps. By utilizing transfer learning in the evaluation of Right Anterior Oblique (RAO) angiograms, we demonstrated the generalizability of image features between different angiographic views.\vspace{-1ex}

\section{Related work}
\IEEEPARstart{D}{eep} learning is a powerful family of data-driven techniques for computer vision, and has been successfully implemented in several medical image domains. Convolutional neural networks (CNN) are a particular type of deep learning that have achieved state-of-the-art accuracy in image classification \cite{NIPS2012_4824}, with several important design innovations such as increasing network depth (VGG) \cite{Simonyan2014VeryDC}, adding shortcut connections between layers (ResNet) \cite{DBLP:journals/corr/HeZRS15} and concatenating outputs from layers (DenseNet) \cite{DBLP:journals/corr/HuangLW16a}. Specifically, the U-Net, in which skip connections are added between layers, achieves a high precision in partitioning images into meaningful components, a task known as image segmentation \cite{DBLP:journals/corr/RonnebergerFB15}. Many studies of automated angiography analysis have employed deep learning techniques, described below.

Ongoing efforts to automate angiography video analysis can be categorized into the following tasks: key frame extraction, vessel segmentation, and stenosis measurement. For key frame extraction, classical methods include the use of image processing techniques for vessel detection, such as the Frangi filter or other edge detection algorithms \cite{5597684}. These techniques are not data-driven, typically require manual parameter tuning, and show limited generalization to images of lower quality or contrast. On the other hand, \cite{DBLP:journals/corr/abs-1804-10021} achieved high performance with a two-stream summarizer-discriminator CNN.

For vessel segmentation, image processing approaches include shape-and-motion-mapping \cite{beymer_2013} and active contour models \cite{6826027}. Recent work employing deep learning have proposed using a multi-channel CNN \cite{8432384}, a combination of two CNNs processing local and global image patches \cite{NASRESFAHANI2018240}, and a multi-channel U-Net \cite{DBLP:journals/corr/abs-1805-06406}. The use of more complex U-Net architectures with DenseNet, and InceptionResNet feature backbones enables higher accuracy \cite{yang_2019}. Most of these studies achieved a high degree of precision. However, few of them sought to apply their algorithms towards the goal of stenosis measurement.

Stenosis measurement is a crucial clinical measurement. Prior work used spatial-temporal information to track vessel width \cite{6868115} and an improved Measure of Match (MoM) measurement \cite{Wan2018}. Recently, an end-to-end analysis was proposed using three CNNs, each responsible for localizing stenosis, segmentation, and comparison between normal and lesion vessel widths \cite{DBLP:journals/corr/abs-1807-10597}, although single-frame images (not videos) were used, and localization/segmentation accuracies were $\sim$70\%. Instead of providing quantitative measurements, another research employed a CNN enhanced by a self-attention mechanism to localize and classify stenosis types \cite{MOON2021105819}.


\begin{figure*}[!ht]
\centering
\captionsetup[subfigure]{labelformat=simple,labelfont = {bf},font = normalsize,farskip = 0pt,captionskip = 0.1pt}
\fbox{\begin{minipage}{0.05\linewidth}\vspace{-16ex}\textbf{A}\end{minipage}\begin{minipage}{0.86\linewidth}\subfloat{\includegraphics[width=\linewidth]{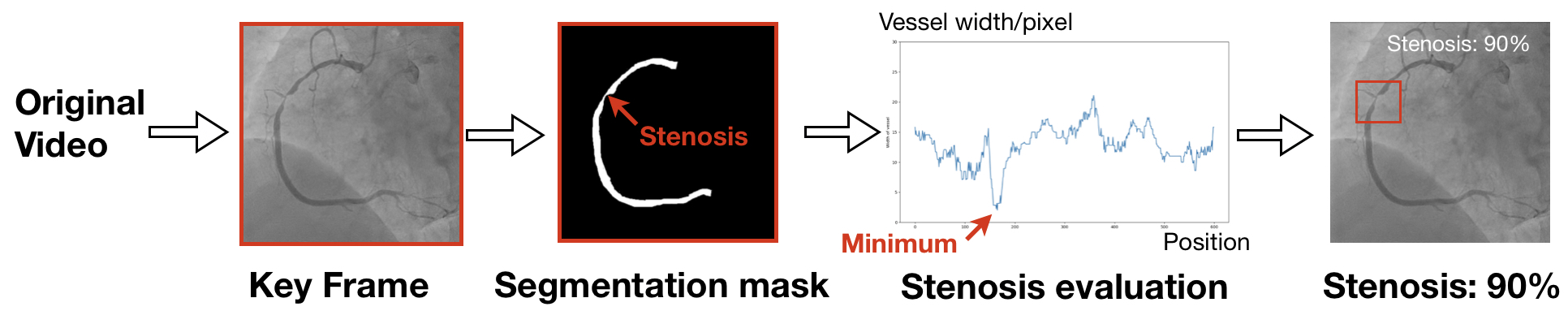}\label{overall-pipeline}}\end{minipage}}
\\
\vspace{0.5ex}
\fbox{\begin{minipage}{0.6\linewidth}\subfloat[Key frame extraction process]{\includegraphics[width = \linewidth]{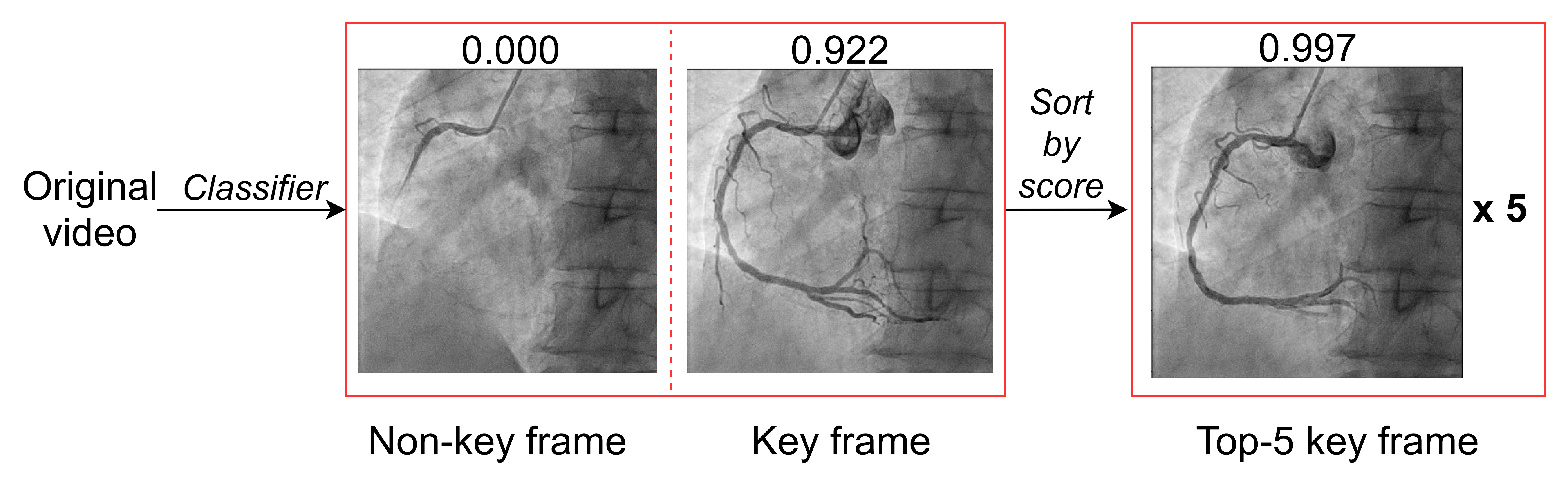}\label{fig:classification-pipeline}}\end{minipage}}\hspace{3mm}
\fbox{\begin{minipage}{0.02\linewidth}\vspace{-20ex}\textbf{C}\end{minipage}\begin{minipage}{0.25\linewidth}\subfloat{\includegraphics[width = \linewidth]{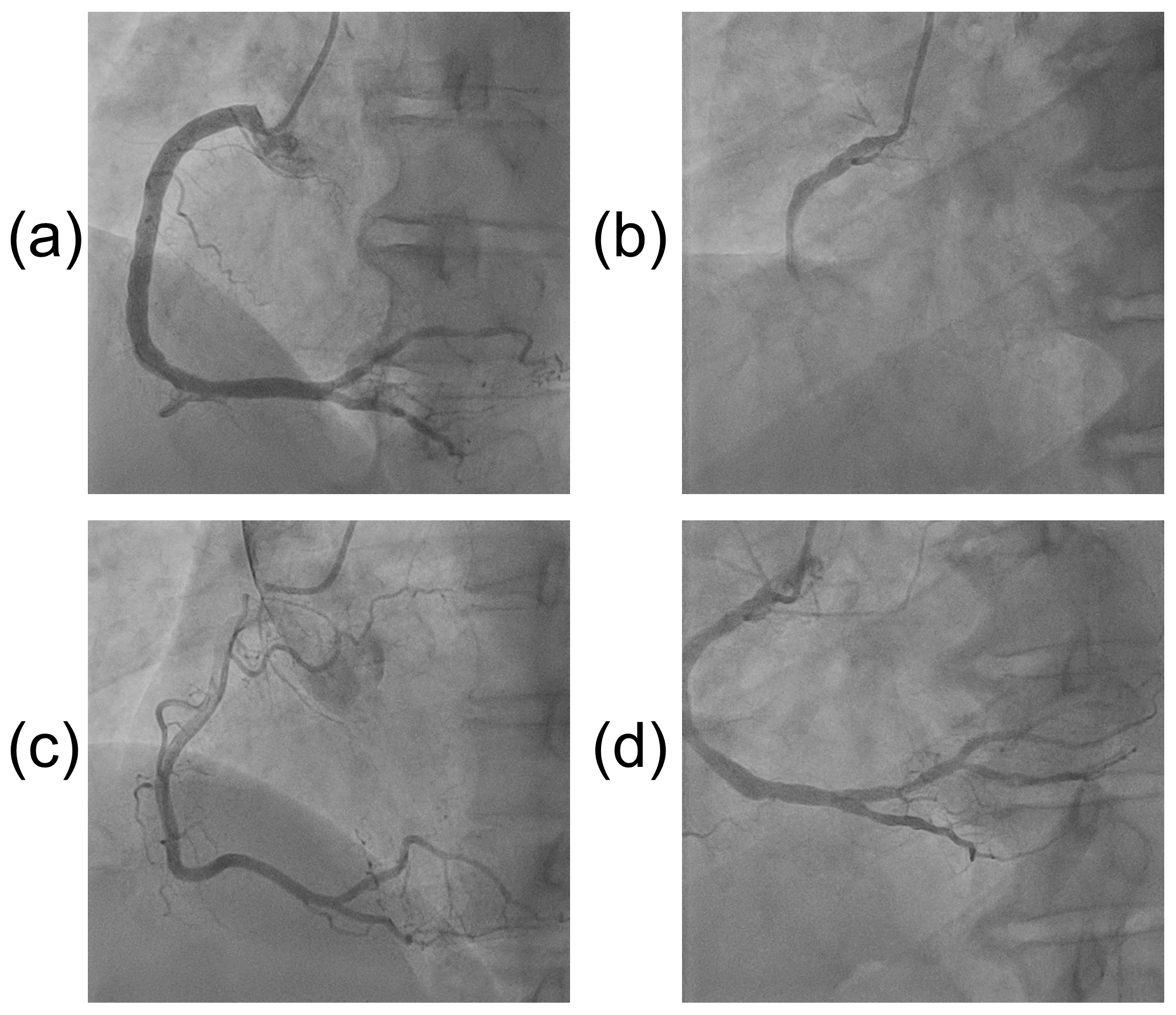}\label{key-frame-labeling}}\end{minipage}}
\\
\vspace{0.5ex}
\fbox{\subfloat[Segmentation labeling]{\includegraphics[width = 0.263\linewidth]{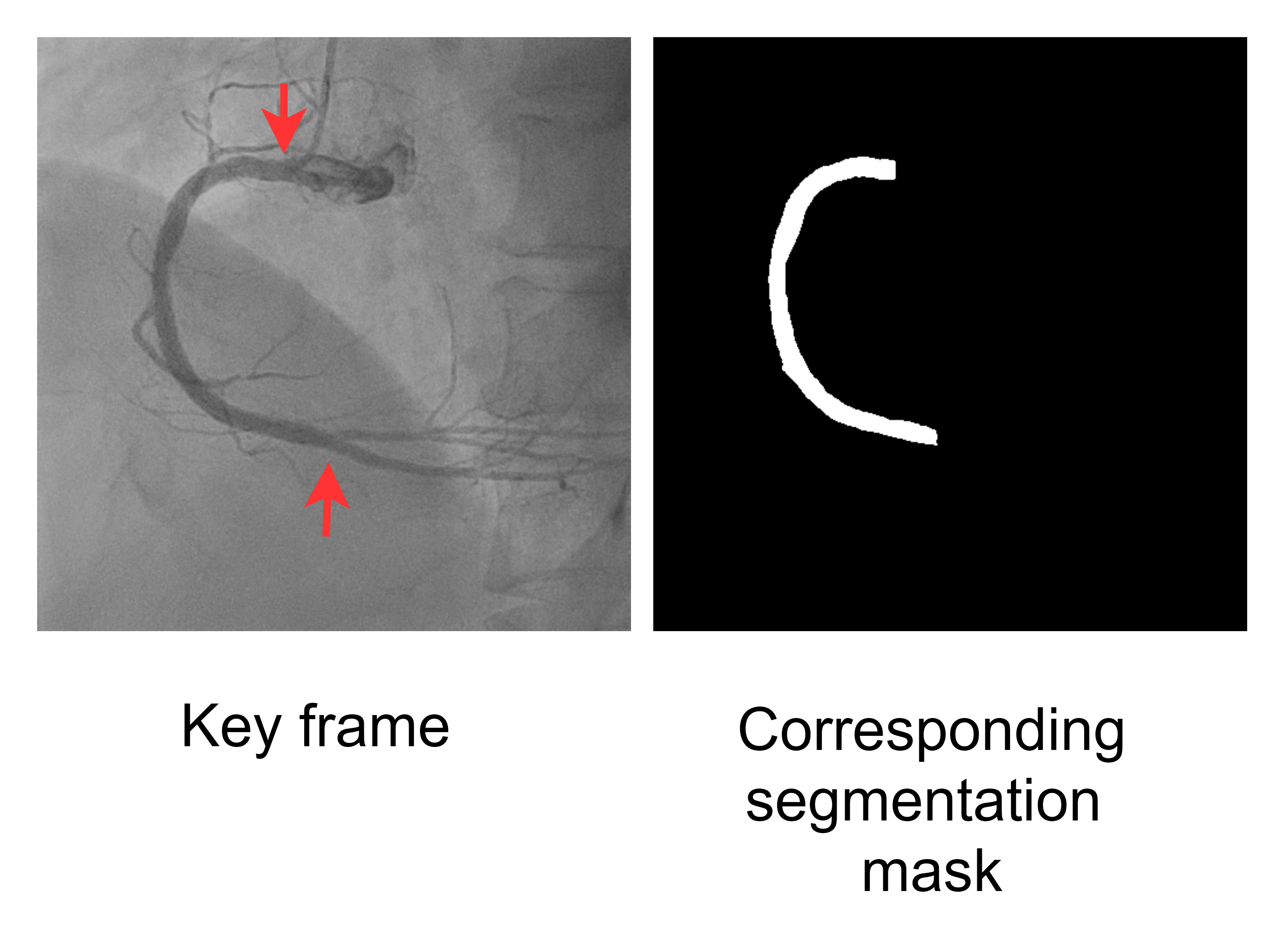}\label{fig:segmentation-label}}}\hspace{1mm}
\fbox{\subfloat[Stenosis measurement algorithm]{\includegraphics[width = 0.62\linewidth]{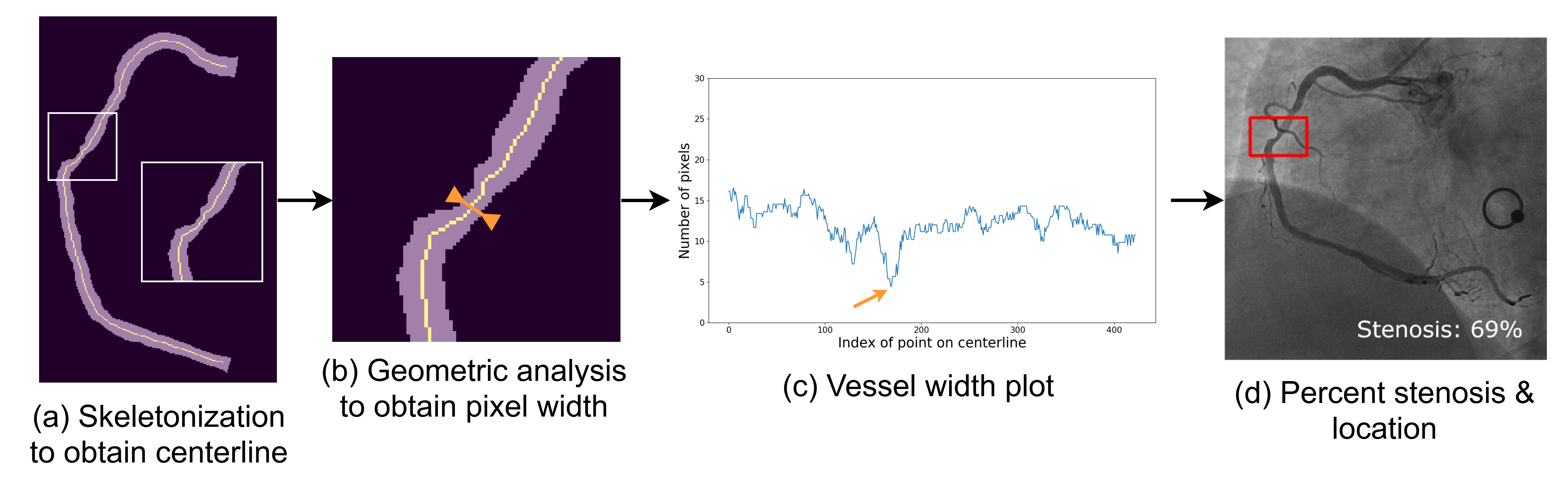}\label{fig:stenosis-pipeline}}}

\caption{Schematics of methodology. (\textbf{A}) Workflow of the proposed angiography analysis pipeline with three main stages: extraction of key frames from video, vessel segmentation of the selected frames, and stenosis measurement through calculating vessel width, giving stenosis location and severity.
(\textbf{B}) Key frame extraction process, showing machine prediction scores, left: two frames from the same video, one non-key (score: 0.000) and one key (score: 0.922), right: a top-5 key frame. 
(\textbf{C})
Illustration of labeled frames. (a) key frame. (b) vessel not fully formed. (c) contrast agent starts fading. (d) vessel has shifted out.
(\textbf{D}) A key frame with corresponding segmentation label (the vessel region between the red arrows is the desired segment for segmentation). (\textbf{E}) Proposed stenosis measurement algorithm. Vessel width along the centerline is plotted to locate the minimum. Note that orange arrow in (c) points to the position of the most severe stenosis.
}
\label{fig: methods}
\vspace{-2ex}
\end{figure*}

\section{Methods}

\IEEEPARstart{W}{e} propose a 3-stage end-to-end pipeline: a) key frame extraction from video sequences, b) vessel segmentation on key frames, and c) stenosis measurement from segmentation masks (Fig. \ref{overall-pipeline}). Stage 1 picks 5 frames with the highest vessel clarity from each video. Stage 2 performs segmentation on the chosen frames to highlight the key vascular structure. Stage 3 identifies the region of highest stenosis and calculates the percent stenosis from the segmented vessel. Our method can potentially provide end-to-end automatic stenosis measurement while allowing stage-by-stage manual interpretation of intermediate results.

\begin{table}[!ht]
\vspace{-1ex}
\centering
\scriptsize
\begin{threeparttable}[t]
\renewcommand{\arraystretch}{1.3}
\caption{Patient demographics and cross-validation folds information}
\label{table:dtu_lao}
\centering

\begin{tabular}{lc lc}

\hhline{====}
 & No. of patients & & No. of patients \\
\hline
\textbf{Total} & \textbf{102} & & \\
\hline
\textbf{Gender (n=90)} & & \textbf{Risk factors (n=90)}\tnote{*} & \\
Female & 23 & Diabetes & 29 \\
Male & 67 & Dyslipidemia & 55 \\
\cline{1-2}
\textbf{Age (n=90)} & & Hypertension & 50 \\
18-39 & 1 & Renal Impairment & 9 \\
40-59 & 34 & Smoking (former) & 16  \\
60-79 & 53 &  Smoking (current) & 12 \\
80+ & 2 & None of the above & 25 \\
\end{tabular}
\setlength{\tabcolsep}{1.5em}
\begin{tabular}{c c c c c}
\hhline{=====}
& Train & Val & Test & Total\\
\hline
No. of patients & 61 & 20 & 21 & 102\\
No. of key frames & 1329 & 397 & 472 & 2198\\
No. of non-key frames & 3927 & 1298 & 1308 & 6533\\
\hhline{=====}

\end{tabular}
\begin{tablenotes}
\item [*] Patients may be subjected to two or more risk factors.
\item [**] Cross-validation fold 1 information shown and other folds follow a similar ratio. 
\end{tablenotes}
\end{threeparttable}

\vspace{-5ex}
\end{table}

\subsection{Dataset}
The coronary arteries consist of the right and left coronary artery. During angiography, images of the coronary arteries are taken at various angular projections, providing comprehensive visualization of the vessels and assessment of stenosis. In this study, we focused primarily on the right coronary artery (RCA) taken in two projections -  the left and right anterior oblique (LAO and RAO). The dataset was obtained from a tertiary cardiac institution. The de-identified images of consecutive patients who underwent coronary angiography for assessment of their cardiac status were included. A summary of demographics is presented in Table \ref{table:dtu_lao}. Videos that had low contrast or visibility of vessels, or had 100\% stenosis causing a substantial alteration in vessel shape were excluded. We employed a set of LAO videos for training and testing of the deep learning models, and a smaller set of RAO videos for testing. After excluding videos of exceptionally low quality, the LAO dataset comprised 102 videos with 8731 frames of size 512x512 pixels, with an average of $\SI{86 \pm 22}{}$ frames per video.\vspace{-1ex}

\subsection{Key Frame Extraction}
Key frame extraction was implemented to select high-quality frames displaying complete vessel shapes for further analysis. We employed a two-phase algorithm for the training of key frame classification models (Fig. \ref{fig:classification-pipeline}). We first trained a base model on resized 64x64 images using manually generated ground truth labels of 6533 non-key and 2198 key frames. The set of criteria for key frames consists of high vessel clarity, a clear vessel edge, and visibility of the proximal, mid and distal RCA i.e. the contrast agent has reached the crux where the RCA divides into the posterior descending and acute marginal arteries (Fig. \ref{key-frame-labeling}). Key frames were labeled by student research assistants (C.Z., T.V.D., H.K.) trained and supervised by an interventional cardiologist (J.Y.) and postdoctoral researcher (K.L.). 

The 102 patients are split into training set and test set based on a 4:1 ratio (81 in training set and 21 in test set). The training set is further divided into 4 equal folds for cross-validation. With patient-level splitting, every non-test set patient appears exactly once in the validation set in order to prevent information leakage between training and validation sets. 
The model runs 500 iterations per epoch for a total of 100 epochs. In each iteration, 16 key and 64 non-key frames are processed, with each key and non-key frame augmented 7-fold and 1-fold in order to mitigate the small and unbalanced nature of our dataset (Training details see Appendix \ref{appendix:training}). Upon completion of training, the classifier is tested on an unseen test set. 

The base model was designed as ResNet18, using implementation from \cite{chollet2015keras}, which employs neural network shortcut paths for better learning \cite{DBLP:journals/corr/HeZRS15}. Dropout layers were used to mitigate overfitting. Two convolutional layers were then added to the base model and the new model was trained on images resized to 128x128 pixels while reusing the weights of the previous phase. This method, known as progressive resizing \cite{bilogur_2019}, uses the base model as a global feature extractor, while the second phase preserves local information by upsizing and training on images with higher resolution. This technique also reduces the computational power required to directly train on high-resolution images. We employed a combination of neural network optimizers: RAdam, an improved Adam \cite{Kingma2014AdamAM} algorithm which stabilizes training with learning rate warmup \cite{Liu2019OnTV}, and Lookahead, which reduces optimizer variance \cite{DBLP:journals/corr/abs-1907-08610}.

Given a single image for prediction, the model generated a score indicating the frame quality (0 - low, 1 - high). Our algorithm then selected 5 frames with the highest scores from each video sequence. In addition to a typical per-frame accuracy metric, we introduced another metric calculating the percentage of true key frames out of the best 5 (``Top-5") output frames of each video sequence (hereafter referred to as the ``Top-5 Precision").\vspace{-2ex}

\subsection{Vessel Segmentation}

After obtaining high-quality key frames, we applied vessel segmentation on them in order to extract clear vessel shapes out of the possibly noisy backgrounds.
For vessel segmentation, we implemented a U-Net model \cite{UnetRepo}. U-Net \cite{DBLP:journals/corr/RonnebergerFB15} is widely used in biomedical image segmentation due to its capability to preserve both global structure and local details; its shortcut connections between contracting and expanding paths facilitate feature reconstruction. Our model consisted of 5 convolutional blocks, followed by a symmetric set of 5 up-convolution blocks. Shortcuts by concatenation were introduced between each convolutional block and its corresponding up-convolution block with the same number of channels. The network took a 512x512 pixel image as input and produced a 512x512 pixel segmentation mask. The U-Net segmented the main part of the vessel i.e. the segment between the catheter-vessel junction and the bifurcation point of the RCA (Fig. \ref{fig:segmentation-label}).

Due to the time-consuming nature of producing precise pixel-level labels for learning segmentation, we implemented semi-supervised label propagation \cite{DBLP:journals/corr/abs-1904-04717}. A small subset (20\%) of the whole dataset was manually labeled and a preliminary U-Net was trained using the subset. The U-Net was then used to generate approximate labels for another 20\% of the dataset. The vessel masks obtained were manually and efficiently corrected, then added to the training data for a subsequent U-Net, thus further improving the approximate labels. This cycle was repeated for two more times. This procedure was used to generate segmentation labels for 2198 key frames, greatly reducing the amount of manual labor required. Data augmentation was performed to mitigate our limited data. Similar to key frame classification, U-Net training was conducted using 4-fold cross-validation, using the same train/validation/test patient cross-validation folds as before. (Training details see Appendix \ref{appendix:training})\vspace{-2.5ex}

\subsection{Stenosis measurement}
To assess the severity of stenosis, physicians typically estimate the degree of narrowing by observing variations in vessel width. Hence, we introduced a stenosis measurement algorithm as the final stage of our end-to-end pipeline.  
We adopted a classical approach combining image processing and geometrical analysis of the segmentation mask. First, the segmentation mask was skeletonized \cite{Zhang:1984:FPA:357994.358023,scikit-image} and pruned (Fig. \ref{fig:stenosis-pipeline} a-b) (Appendix \ref{appendix:algorithms}, Algorithm \ref{alg:prune}) to produce a centerline. Subsequently, the width of the vessel was approximated using the geometry of the vessel slope at multiple points along the smoothed centerline (Fig. \ref{fig:stenosis-pipeline} c-d). This type of measurement has been previously proposed for direct use on angiograms\cite{bove_estimation_1985}; we expected it to perform more reliably on a binary mask, which has clear edge boundaries. This vessel width estimation algorithm (Appendix \ref{appendix:algorithms}, Algorithm \ref{alg:vessel-width-plot}), was repeated for all top-5 key frames from the same video. An estimation of percent stenosis was the quotient of two quantities: the ``normal" (non-pathological) width of the vessel, and the minimum (pathological) width. We estimated the normal width by taking the average of the 30 largest point approximations of vessel width per frame. The minimum width was estimated by taking the average of the 3 lowest pixel widths, assuming that the stenosis was of a focal nature. The stenosis severity was then calculated as follows: percent stenosis = $(1- \frac{\text{average minimum width}}{\text{average maximum width}}) \times 100\%$.

\begin{table*}[t]
\vspace{-2ex}
\centering
\scriptsize
\begin{threeparttable}[t]
\renewcommand{\arraystretch}{1.3}
\caption{Model performance on key frame extraction, vessel segmentation and stenosis measurement}
\label{table:all-results}
\centering
\begin{tabular}{C{1.0cm}C{0.7cm}C{0.7cm}C{0.9cm}C{0.9cm}C{0.7cm}C{0.7cm}C{0.05cm}C{0.7cm}C{0.7cm}C{0.05cm}C{1.7cm}C{1.3cm}C{1.3cm}}
\hhline{==============}
\multirow{3}{*}{Fold No.} & \multicolumn{6}{c}{\makecell{Key Frame Extraction}} & & \multicolumn{2}{c}{\makecell{Vessel Segmentation}} & & \multicolumn{3}{c}{\makecell{Stenosis Measurement}} \\ 
\cline{2-7} \cline{9-10} \cline{12-14} & \multicolumn{2}{c}{\makecell{Acc (\%)}} & \multicolumn{2}{c}{\makecell{Top-5 Precision (\%)}} & \multicolumn{2}{c}{\makecell{F1-Score}} & & \multicolumn{2}{c}{\makecell{F1-Score}} & & MAE ${\pm}$ SD of AE\tnote{1} & Acc. for severe lesion (\%)\tnote{2} & Acc. for moderate lesion (\%)\tnote{3} \\ 
\cline{2-7} \cline{9-10} \cline{12-14}
& Val & Test & Val & Test & Val & Test & & Val & Test & & \textit{(n=48)\tnote{*}} & \textit{(n=48)} & \textit{(n=48)} \\
\hline
1 & 91.4 & 88.3 & 99.0 & 97.1 & 0.834 & 0.793 & & 0.913 & 0.887 & & \multirow{5}{*}{\parbox{1.1cm}{\centering \textbf{15.9\%} $\pm$ \\ 13.3\%}} & \multirow{5}{*}{\textbf{79.2}} & \multirow{5}{*}{\textbf{81.3}} \\ 
2 & 90.8 & 90.4 & 96.0 & 98.1 & 0.840 & 0.836 & & 0.871 & 0.890 & & & & \\
3 & 89.2 & 89.6 & 96.2 & 99.1 & 0.774 & 0.810 & & 0.887 & 0.895 & & & & \\
4 & 89.9 & 89.3 & 98.0 & 99.1 & 0.815 & 0.810 & & 0.905 & 0.892 & & & & \\
\textbf{Avg.} & \textbf{90.3} & \textbf{89.3} & \textbf{97.3} &
\textbf{98.4} & \textbf{0.816} & \textbf{0.812} &&  \textbf{0.894} & \textbf{0.891} & & & & \\
\hhline{==============}
\end{tabular}
\begin{tablenotes}
\item [*] Evaluation results are obtained from the only 48 videos  with available clinical assessments of the percent stenosis. These videos are sourced from both the validation sets and the hold-out test set. Note that there is no cross-validation at this stage.
\item [1] Mean Absolute Error ${\pm}$ Standard Deviation of Absolute Error.
\item [2, 3] Accuracy by setting 70\% i.e. severe lesion \& 50\% i.e. moderate lesion as the binarizing thresholds.
\end{tablenotes}
\end{threeparttable}
\vspace{-2ex}
\end{table*}

\subsection{Extension to Right Anterior Oblique (RAO) View}
\begin{figure}[t]
\vspace{-1ex}
    \centering
    \includegraphics[width=1.0\linewidth]{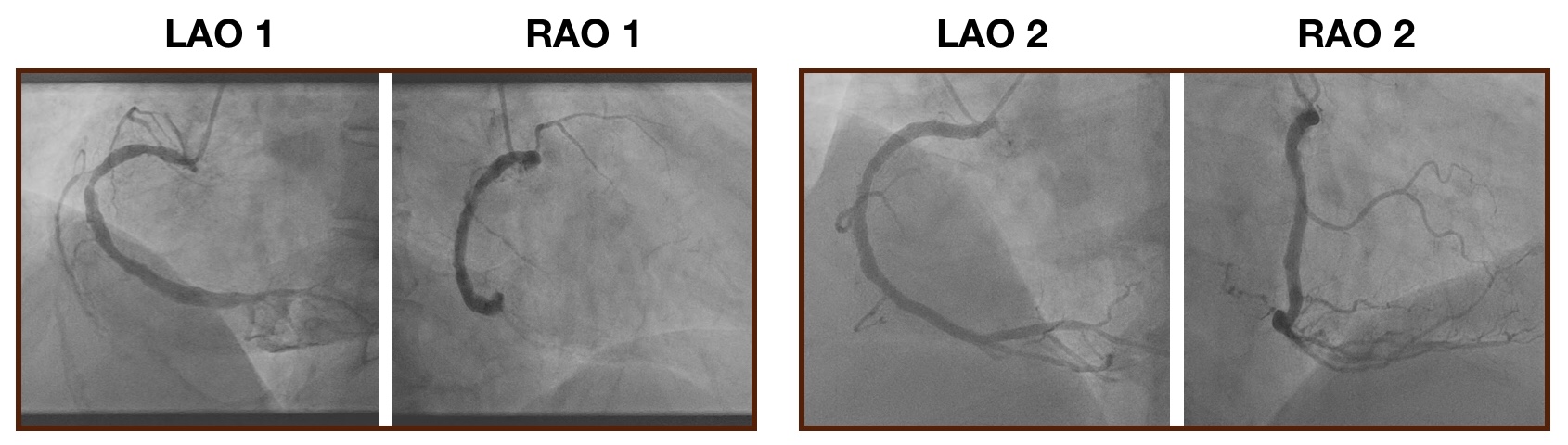}
    \caption{Side-by-side comparisons between LAO and RAO angiograms. Each LAO-RAO pair represents one patient.}
    \label{fig:13}
    \vspace{-3ex}
\end{figure}

Our three-stage pipeline is trained on data from one specific angular projection of a set of up to 9 projections in cardiology practice, namely the LAO straight projection of the right coronary artery. The Right Anterior Oblique (RAO) projection of the RCA is taken at approximately right angles to the LAO, such that the vessel is the same but has a slightly different shape and appearance due to the orthogonal view projection. Fig. \ref{fig:13} shows examples of side-by-side comparisons between the two views. Given the similarities in vessel anatomy, we tested the ability of our trained classification and segmentation models that had only previously seen LAO data to also interpret RAO data. Through this extension, we examined our algorithms' capability to generalize on unseen but related data. We used a test set of 17 RAO videos with 1500 frames (1038 non-key and 462 key). Subsequent results are generated using the test set. In the first two stages of key frame extraction and vessel segmentation, the RAO videos were processed using the trained LAO model weights. For the purpose of evaluating the generalizability of our model, no additional model training on RAO data was done. In the third stage, the same stenosis measurement algorithm was used. \vspace{-2ex}

\section{Results \& Discussion}

\IEEEPARstart{T}{he} experiments for each stage are conducted in independent settings. Specifically, the key frames used in stage 2 and the segmentation masks used in stage 3 are ground truth labels. This is to serve as a feasibility test (proof of concept) of our proposal.\vspace{-1ex}

\subsection{Key Frame Extraction}

Table \ref{table:all-results} shows the per-frame analysis performance of our model, and the per-video Top-5 Precision metric. On a per-frame basis, we factor in the unbalanced nature of our dataset (6533 non-key vs 2198 key) by including F1-score as a metric. Moreover, in practice it is not necessary for every key frame (often up to 20 available in each video) to be correctly identified; hence the Top-5 metric is intended as a more realistic measure of whether an adequate number of key frames (5 as a convenient threshold) may be identified per video. 

\begin{table}[t]
\vspace{-1ex}
\centering
\scriptsize
\begin{threeparttable}[t]
\caption{Comparison with related work}
\renewcommand{\arraystretch}{1.3}
\centering
\begin{tabular}{cC{0.1cm}c}

\hhline{===}
\multicolumn{3}{c}{\makecell{Vessel Segmentation}}\\
\hline
Method\tnote{1} & & F1-score\\
\cline{1-1} \cline{3-3}
Yang et al.\cite{yang_2019} & & 0.930\\ 
\textbf{Our work} & & \textbf{0.891} \\

\hhline{===}
\multicolumn{3}{c}{\makecell{Stenosis Measurement}}\\
\hline
Method & & Type I Error (\%) \\
\cline{1-1} \cline{3-3}
MWCE-End-to-End\tnote{2}\cite{DBLP:journals/corr/abs-1807-10597} & & 36.8 \\
12-feature classifier\cite{Cho2019} & & 16.2 \\
\textbf{Our work} & & \textbf{20.7} \\
\hhline{===}

\end{tabular}
\begin{tablenotes}
\item [1] Significant difference in dataset size exists (1021 patients for Yang et al. and 102 patients for our work).
\item [2] MWCE-End-to-End model's workflow is subjected to potential errors during the former stage.
\end{tablenotes}

\label{table:comparison table}
\end{threeparttable}
\vspace{-3.5ex}
\end{table}

We observe that the metric adopted by previous work such as Frangi Filter and Edge Detection\cite{5597684} is different (the percentage of videos whose output contains at least 1 key frame) from top-5 precision, and is hence not a suitable basis for comparison. The apparent improvement demonstrated by our top-5 precision illustrates the robustness of data-driven approaches such as deep learning. 

\begin{figure*}[!ht]
\vspace{-1ex}

    \centering
    \captionsetup[subfigure]{labelformat=empty}
    \fbox{\begin{minipage}{0.03\linewidth}\vspace{-28ex}\textbf{A}\end{minipage} \begin{minipage}{0.43\linewidth}\subfloat{\includegraphics[width = \linewidth]{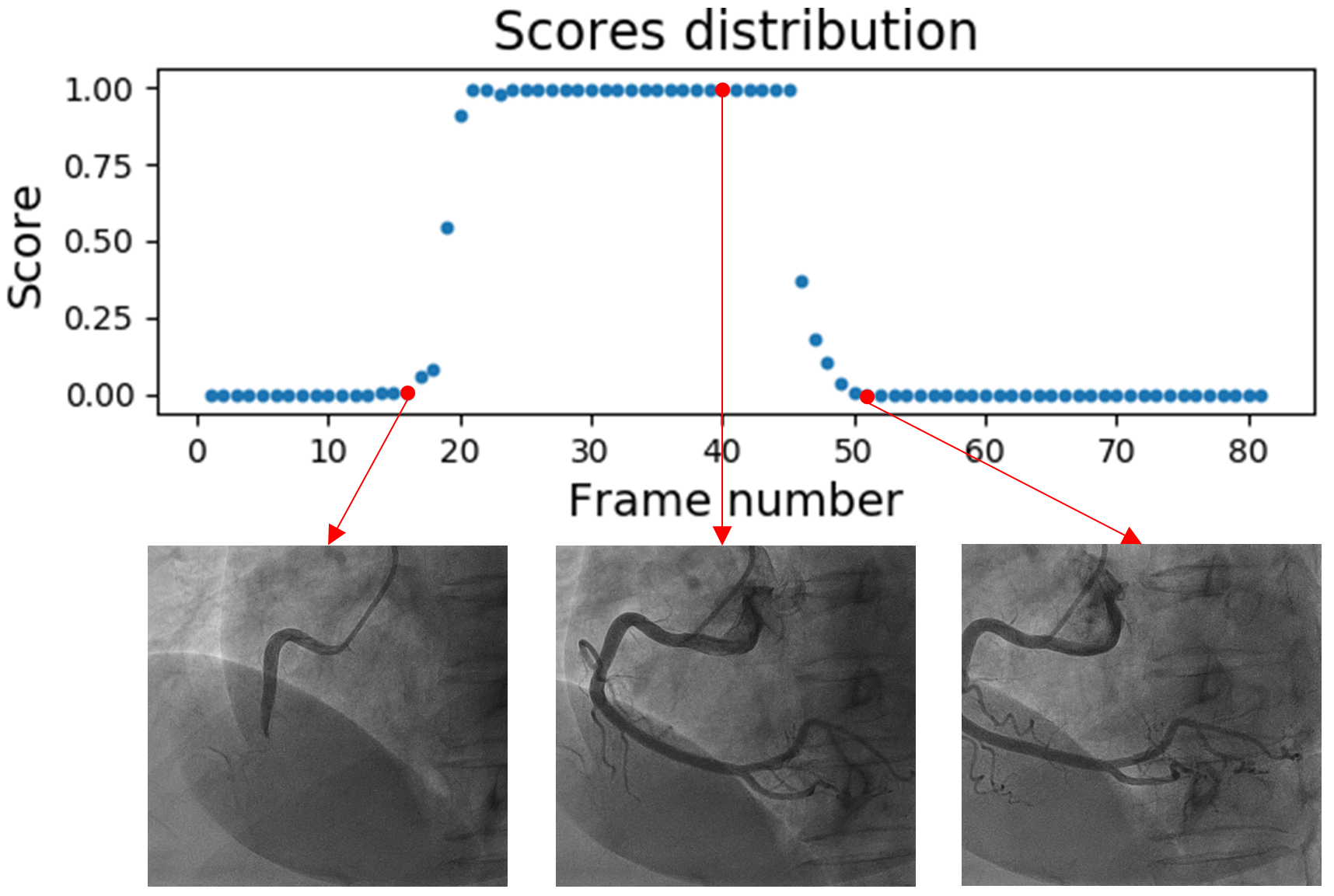}\label{fig:scores-distribution}}\end{minipage}}\hspace{1mm}
    \fbox{\begin{minipage}{0.03\linewidth}\vspace{-28ex}\textbf{B}\end{minipage}\begin{minipage}{0.43\linewidth}\subfloat{\includegraphics[width = \linewidth]{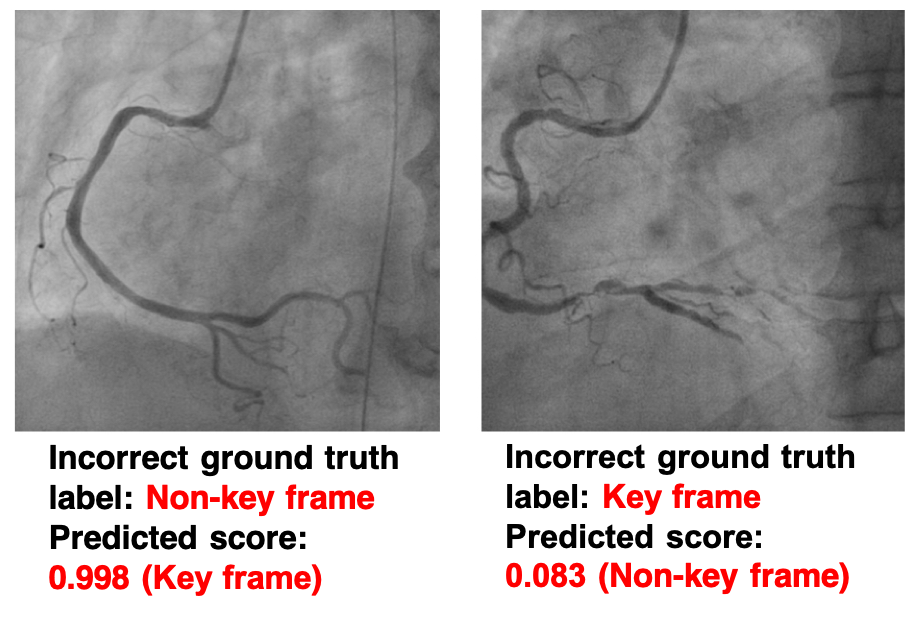}\label{fig:Example of machine's inferential ability}}\end{minipage}}
    \\
    \vspace{0.5ex}
    \fbox{\begin{minipage}{0.04\linewidth}\vspace{-26ex}\textbf{C}\end{minipage}\begin{minipage}{0.42\linewidth}\subfloat{\includegraphics[width = \linewidth]{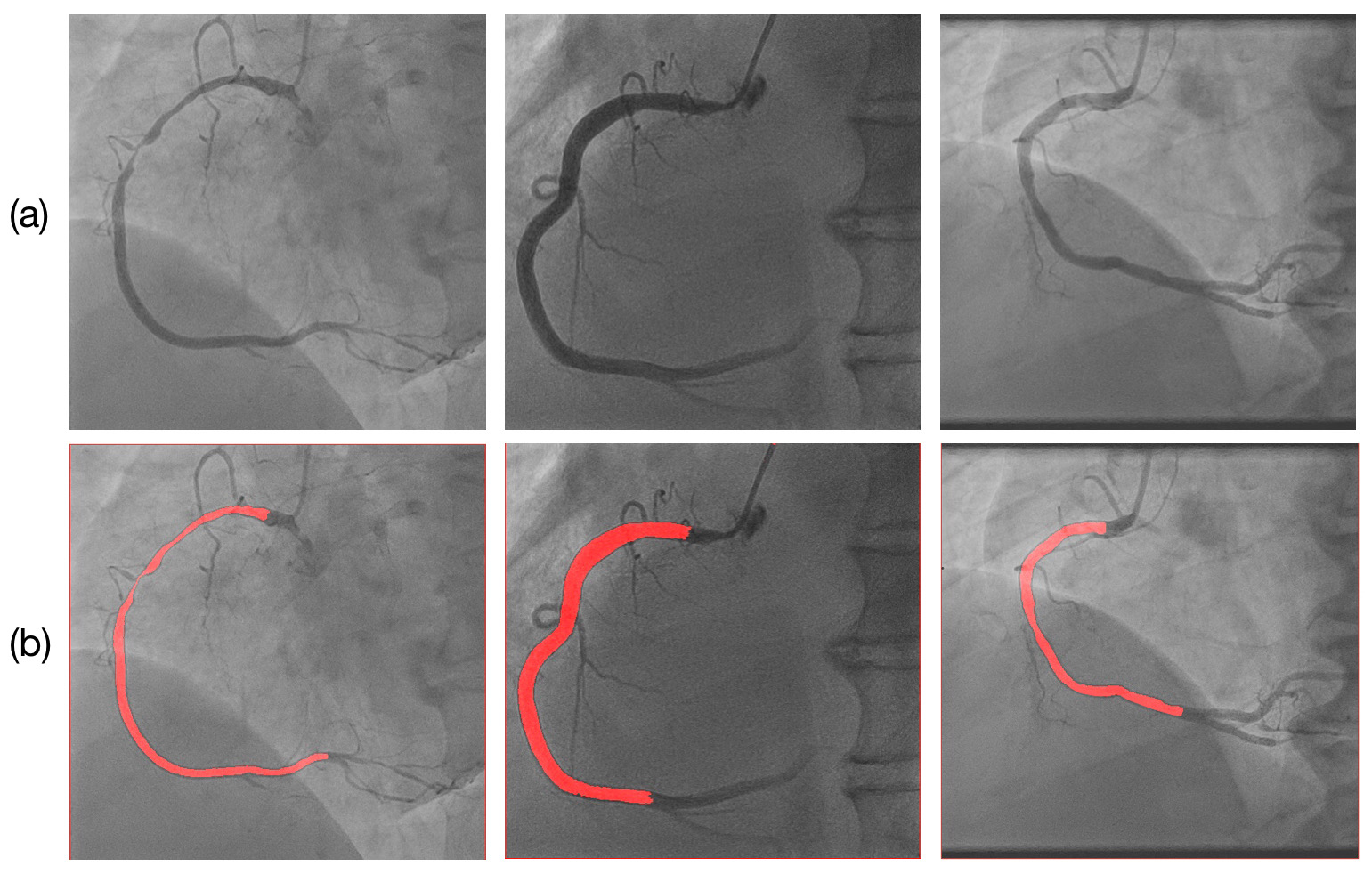}\label{fig:LAO-segmentation}}\end{minipage}}\hspace{1mm}
    \fbox{\begin{minipage}{0.02\linewidth}\vspace{-26ex}\textbf{D}\end{minipage}\begin{minipage}{0.45\linewidth}\subfloat{\includegraphics[width=\linewidth]{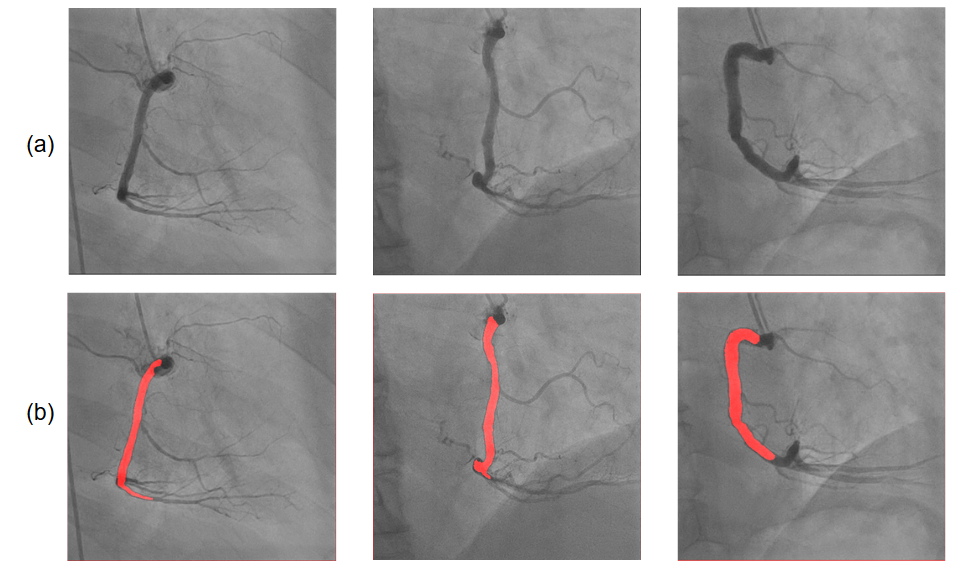}\label{fig:RAO-segmented-masks}}\end{minipage}}
    \\
    \vspace{0.5ex}
    \fbox{\begin{minipage}{0.02\linewidth}\vspace{-15ex}\textbf{E}\end{minipage} \begin{minipage}{0.44\linewidth}\subfloat{\includegraphics[width = \linewidth]{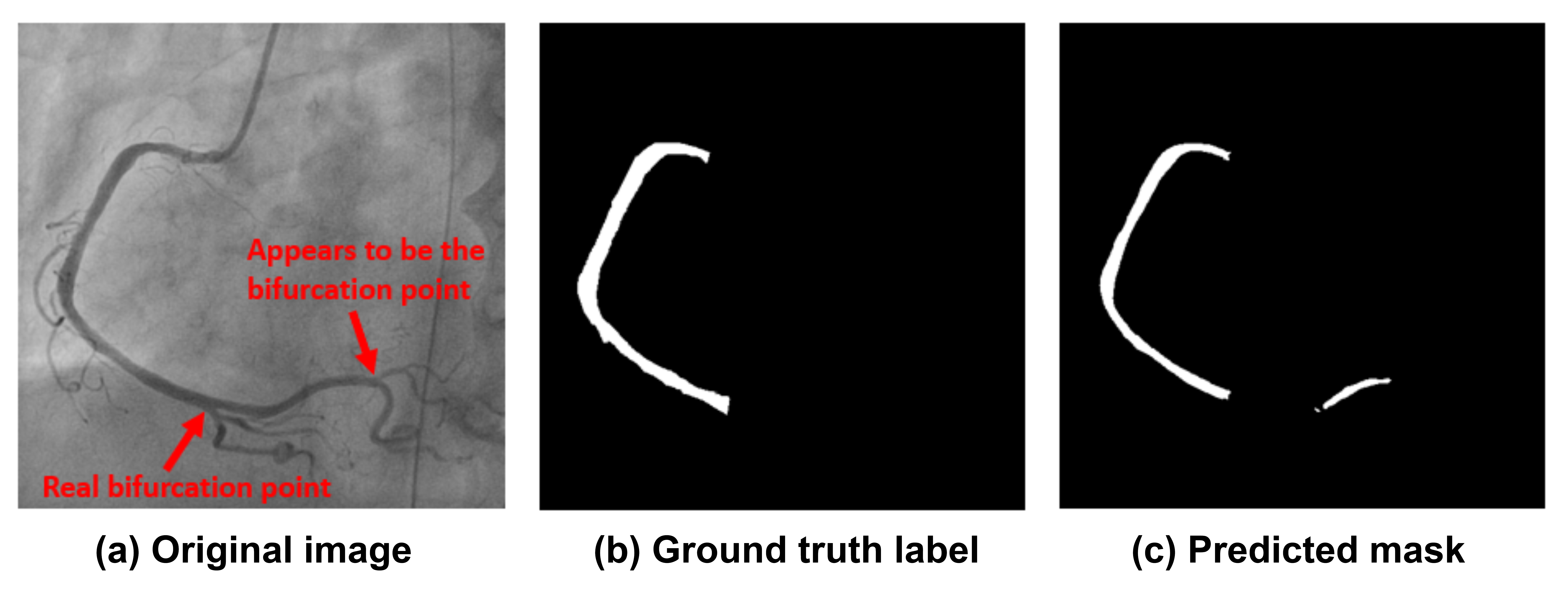}\label{fig:segmentation-failure}}\end{minipage}}\hspace{1mm}
    \fbox{\begin{minipage}{0.015\linewidth}\vspace{-15ex}\textbf{F}\end{minipage}\begin{minipage}{0.45\linewidth}\subfloat{\includegraphics[width = \linewidth]{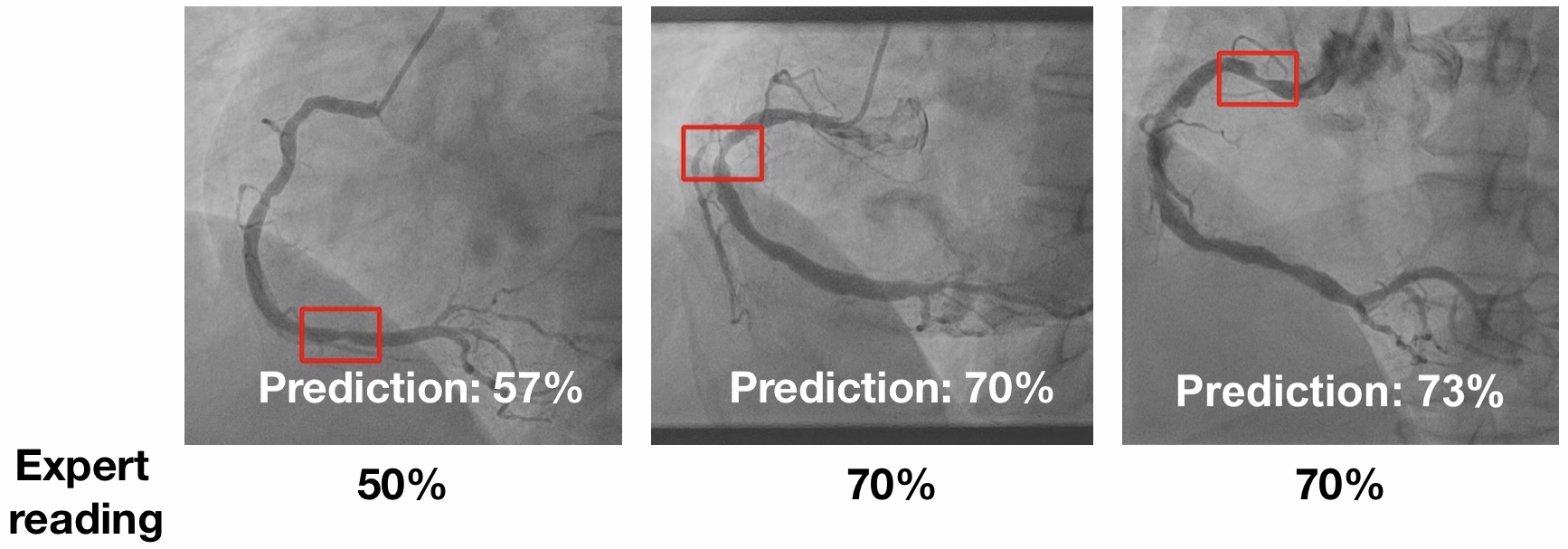}\label{fig:LAO-stenosis}}\end{minipage}}
    \\
    \vspace{0.5ex}
    \fbox{\begin{minipage}{0.04\linewidth}\vspace{-31ex}\textbf{G}\end{minipage}\begin{minipage}{0.92\linewidth}\subfloat{\includegraphics[width=\linewidth]{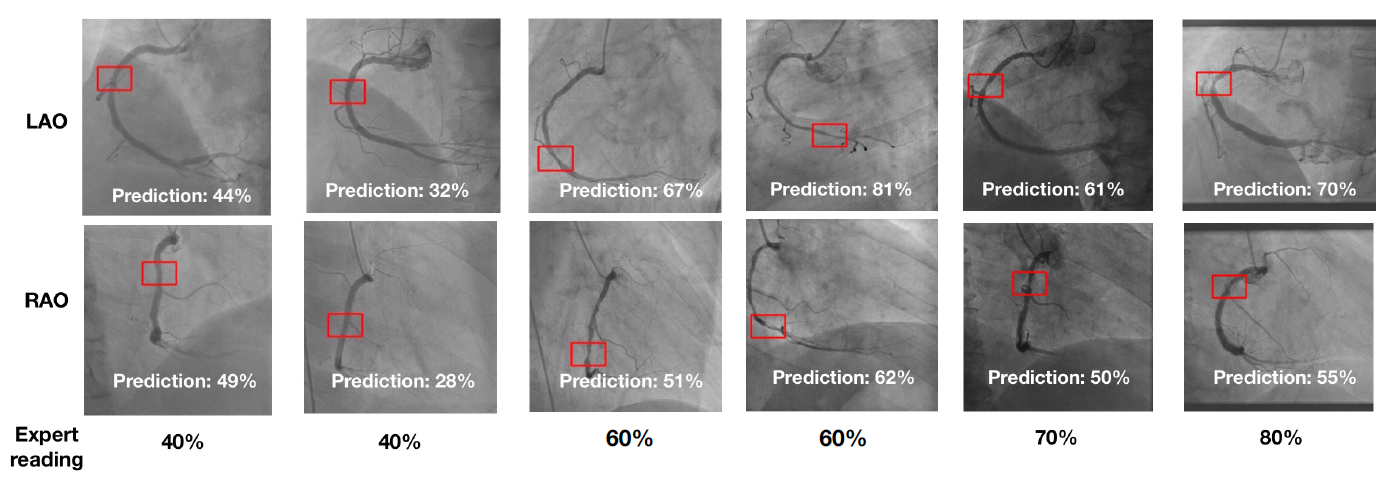}\label{fig:LAO-RAO-stenosis-side-by-side}}\end{minipage}}
    
    \caption{Illustration of LAO and RAO results. (\textbf{A}) Temporal score variation over the course of an LAO angiography video. Each data point represents the machine predicted score of that frame. Three particular frames are highlighted in red as they represent the frame just before, in and after the ``key frame region". (\textbf{B}) Machine's inferential ability shown by noisy incorrect LAO labels detected by the model. First image: a  key  frame  erroneously  labeled  as  non-key  in  manual labeling, and the model gave the frame a high score (0.998) Second image: a non-key frame with vessel shifted out erroneously labeled as a key frame;   it was detected  and  given  a  low  score (0.083) (\textbf{C}) LAO vessel segmentation, (a) original images, (b) segmentation masks.
    (\textbf{D}) RAO vessel segmentation, (a) original images, (b) segmentation masks.  (\textbf{E}) Erroneous LAO segmentation near the RCA bifurcation point.  (\textbf{F}) Identification of LAO stenosis location and estimation of severity. (\textbf{G}) Side-by-side comparisons between LAO and RAO stenosis measurement. Each column represents one patient.}
    \label{fig: results}
    \vspace{-3.5ex}
\end{figure*}

The model performance was partly hindered by ``noisy" ground truth labels due to inter-observer variations among our human readers, especially on frames that occurred on the margin of the contrast-enhanced time period in each video. We observed that some of these errors could be ``corrected" by the trained model; a key frame erroneously labeled as non-key in manual labeling could be detected by the model, giving the frame a high score; similarly, a non-key frame with incomplete vessel shape erroneously labeled as key frame  was detected and given a low score (Fig. \ref{fig:Example of machine's inferential ability}).

Since the algorithm was trained on individual frames and analyzed frames one at a time, we did not incorporate temporal information that may provide additional information to the machine learning; alternate designs of convolutional networks incorporating time-series data could be investigated in the future. Nevertheless, our per-frame analysis could reconstruct the temporal relationship within video sequences, where predicted scores showed a characteristic curve (Fig. \ref{fig:scores-distribution}), as key frames typically cluster around the middle of a video (referred to as the ``key frame region"). Frames just before, in, and right after the key frame region are highlighted, illustrating the sensitivity of the model to changes in the visibility and contrast of the vessel anatomy.\vspace{-1ex}

\subsection{Vessel Segmentation}

Table \ref{table:all-results} shows the results of our model, with F1-score as the metric, given by: 
\begin{equation}
    \text{F1-score} = \frac{2\sum_{i=1..512} \prediction \times \truelabel}{\sum_{i=1..512} \prediction + \sum_{i=1..512}\truelabel}
\end{equation}
where $\prediction$ and $\truelabel$ are arrays representing generated and ground-truth masks respectively. Our segmentation model can localize heart vessels with relatively high precision (Fig. \ref{fig:LAO-segmentation}). Comparing the performance of our model with Yang et al. \cite{yang_2019}, our model using a similar but much smaller training dataset (102 vs 1021 patients) was able to reach a comparable result (Table. \ref{table:comparison table}). Our neural network architecture was also of lower complexity, requiring less computational power for training, albeit prone to errors especially near the distal RCA bifurcation (Fig. \ref{fig:segmentation-failure}). More complex U-Net architectures and more training data will improve the performance in these difficult areas.\vspace{-2ex}

\subsection{Stenosis Measurement}

In predicting the stenosis, our algorithm provides the percent stenosis and approximate position of the most severe obstruction (Fig. \ref{fig:LAO-stenosis}). The performance of the algorithm was evaluated by comparing its predictions to cardiologists' assessments (Table \ref{table:all-results}). The dataset for evaluation consists of 48 videos with available clinical assessments of the percent stenosis obtained from patients' reports. Here, we did not make a distinction between the training and test sets due to the non-data-driven nature of the algorithm, which means it operates in a ``blind" manner. Quantitative metrics include Mean Absolute Error (MAE) and Standard Deviation of Absolute Error (SD of AE). 
\begin{equation}
    \text{MAE} = \frac{\sum_{i=1..n} |\truevalue - \prediction|}{\n}
\end{equation}
where $\truevalue$ and $\prediction$ represent the true clinical assessment of the percent stenosis and machine prediction of the percent stenosis respectively, and $\n$ represents the total number of patients. As part of additional qualitative analysis, we also obtained the accuracy for detecting severe and moderate stenosis of the algorithm by binarizing the percent stenosis at the 70\% and 50\% thresholds as per typical clinical practice for significant lesions and moderate lesions respectively \cite{Cheng2013}. Patients whose percent stenosis are classified as severe lesions are highly recommended to undergo interventionist surgical procedures to restore normal blood flow, while those with moderate lesions are rated on a case-by-case basis by cardiologists. Our model's performance surpasses the MWCE-End-to-End model\cite{DBLP:journals/corr/abs-1807-10597} but misses the 12-feature classifier\cite{Cho2019} by a 4.5\% margin (Table \ref{table:comparison table}).


The algorithm relies on simple geometrical relationships to compute percent stenosis, which is intuitive and analogous to human readings. However, its simplicity may have led to a relatively large error margin (15.9\% ${\pm}$ 13.3\%). Firstly, stenoses in the most severe cases such as complete or near-complete (99\%) occlusions were underestimated or missed out due to the algorithm's inability to detect ``invisible" vessels i.e. vessels that the contrast agent was unable to reach. Secondly, the model sometimes performed poorly in identifying lesions at the early proximal and late distal segments due to errors in the skeletonization algorithm. Furthermore, this algorithm considers the vessel's maximum width as the reference width, but other options such as local maxima proximal to lesions are also available and can be more intuitive to cardiologists \cite{Wan2018,Ota2005}. \vspace{-2ex}

\subsection{End-to-end pipeline}

We developed an end-to-end tool in a prototype software interface that integrates our machine models and algorithms. The software takes in an angiography video as input, and performs analysis, after which the user may navigate the results of key frame extraction, vessel segmentation, and stenosis measurement. The software took approximately 30 seconds for an end-to-end analysis. We believe that this class of automated tools can serve to assist cardiologists in their diagnoses, as well as rapidly generate repeatable and objective measurements on large datasets in a high-throughput fashion.

While our goal was to demonstrate the feasibility of such a multi-stage algorithm, we assessed the performance of each of the three analysis stages separately. The segmentation algorithm was trained and evaluated using all key frames, rather than key frames selected by the key frame classifier. Similarly, the stenosis measurement algorithm was evaluated on all ground truth segmentation masks, rather than actual U-Net predictions. This allowed us to objectively evaluate the algorithmic performance of each stage and compare our results to prior studies, but did not reflect true end-to-end performance.\vspace{-2ex}
\subsection{Extension to Right Anterior Oblique View}
\begin{table}[t]
\centering
\scriptsize
\begin{threeparttable}[t]
\renewcommand{\arraystretch}{1.3}
\caption{Comparison between LAO and RAO results}
\label{table: RAO-vs-LAO}
\centering
\begin{tabular}{cC{0.1cm}cC{0.1cm}c}

\hhline{=====}
Metric & & LAO & & \textbf{RAO}\\
\hhline{=====}
\multicolumn{5}{c}{\makecell{Key Frame Extraction}}\\
\hline
Acc (\%) & & 89.3 & & \textbf{84.6} \\ 
Precision & & 0.761 & & \textbf{0.716} \\
Recall & & 0.873 & & \textbf{0.834} \\
Top-5 Precision (\%) & & 98.4 & & \textbf{94.7} \\

\hhline{=====}
\multicolumn{5}{c}{\makecell{Vessel Segmentation}}\\
\hline
F1-Score & & 0.891 & & \textbf{0.826} \\

\hhline{=====}
\multicolumn{5}{c}{\makecell{Stenosis Measurement}}\\
\hline
MAE ${\pm}$ SD (\%) & & 13.5 ${\pm}$ 10.2 & & \textbf{10.1 ${\pm}$ 8.1} \\
Acc. for severe lesion (\%) & & 71.4 & & \textbf{88.2} \\
Acc. for moderate lesion (\%) & & 95.2 & & \textbf{82.4} \\
\hhline{=====}

\end{tabular}
\end{threeparttable}
\vspace{-2ex}
\end{table}

Key frame extraction and vessel segmentation performance evaluation on LAO and RAO data are presented in Table \ref{table: RAO-vs-LAO}. A slight dip in performance is seen using the RAO view in comparison to LAO, which can be attributed to the fact that the models were intially trained on LAO data. However, results on RAO are remarkably robust, indicating that the trained models have some generalizability to vessels with similar anatomical structures. Segmentation F1-score for the RAO view also did not dip significantly. The predicted masks delineated the RCA vessel structures clearly and included appropriate endpoints at their distal bifurcations (Fig. \ref{fig:RAO-segmented-masks}), which were landmarks learned from LAO labels. Further fine-tuning of the models on a moderate amount of true RAO labels can greatly boost performance if required. 

As a preliminary exploration of how automated measurements of a given lesion from different views may be aggregated, we performed manual visual comparisons between the identified stenosis positions of lesions in LAO and RAO views (Fig. \ref{fig:LAO-RAO-stenosis-side-by-side}). The algorithm could identify the correct location of the most severe blockage in nearly all videos (14/15 patients). The automatically boxed regions highlighting the most severe stenosis in the LAO and RAO views correspond to each other.

The algorithm evaluates the most severe obstruction in a given view; different views highlight different segments of the vessel, and thus this aggregation of observations from multiple views is expected to improve the accuracy of the prediction, and simulating a cardiologist's practice. We suggest using the higher percent stenosis of the two predictions retrieved from the LAO and RAO views i.e. Max(LAO, RAO). To validate this concept, a subset of 15 patients (test set) with both LAO and RAO readings was used. Using Max(LAO, RAO) gave $\SI{12.0 \pm 9.1}{\percent}$ as MAE ${\pm}$ SD of AE. The correct detection rate of the stenosis at the 70\% thresholds was 73.3\%, reaching 100\% at the 50\% threshold.

The extension to the RAO view provides inspiration for further research. Future work include expansion to more views such as the anterior-posterior (AP) Cranial view to visualize the more distal vessel segments. Leveraging information from individual views and combining them in an optimal way will improve our model's accuracy. The ability to integrate information from multiple views will be even more critical as we continue work on the left coronary artery, which has a more complex vessel architecture studied from up to 6 view projections.\vspace{-0.5ex}

\section{Conclusion}

We developed a set of algorithms designed as an integrated end-to-end automated tool for analyzing angiography video sequences. Automating the analysis of coronary angiograms could assist cardiologists with visual assessment and report generation of these angiograms, while mitigating inter-observer variability and preventing unnecessary stenting. The capability to process large datasets in a repeatable, objective fashion could also be an important tool in clinical trials and research. Future improvements include: 1. the addition of video sequences with artefacts and near-complete occlusions (99\%) into the dataset and subsequent model tuning, 2. incorporation of time-series data into the current classification model, and 3. employing a more complex stenosis measurement algorithm combining information from vessel widths proximal to lesions and other local maxima/minima features. Additionally, regression models are a possible alternative to the current non-data-driven approach utilized in stage 3. Extending these concepts to the left coronary artery and more view projections (cranial, caudal), as well as an evaluation of end-to-end performance on larger unseen datasets will be an essential component in realizing the potential for clinical translation. Ultimately, after rigorous validation by experts, these automated tools could have a role in suggesting recommendations for interventional treatment, with even more value to community or rural hospitals that may lack specialist expertise or infrastructure.
\vspace{-1ex}
\section*{Acknowledgment}
This work is supported by the Agency for Science, Technology and Research AI and Analytics Seed Grant no. Z20F3RE003.
\vspace{-1ex}
\bibliography{main.bib}
\bibliographystyle{IEEEtran}

\appendices
\section{Algorithms}
\label{appendix:algorithms}

\begin{algorithm}
\caption{Pruning algorithm}\label{alg:prune}
\begin{flushleft}
\textbf{Input} {Set of points forming the centerline $\mathcal{S}$}
\end{flushleft} 
\begin{algorithmic}[1]
\scriptsize
\Procedure{Prune}{$S$}
    \State $neighbors$ of a point$\gets$ points in $\mathcal{S}$ within the 8-connected neighborhood of the reference point
    \State $end points\gets$ points in $\mathcal{S}$ with 2 $neighbors$
    \For{$endpoint$ in $endpoints$}
        \State{$bifurcationpoint\gets$ point in $\mathcal{S}$ with 4 $neighbors$ that is nearest to $endpoint$}
        \State{$branch\gets$ set of points including $endpoint$, $bifurcationpoint$ and all points in $\mathcal{S}$ that are in between}
        \If{$\vert branch\vert \geq$ 25}
            \State{remove $branch$}
        \EndIf
    \EndFor
\EndProcedure
\end{algorithmic}
\end{algorithm}

\begin{algorithm}
\caption{Overall vessel width extraction algorithm}\label{alg:vessel-width-plot}
\begin{flushleft}
\textbf{Input} {Vessel masks acquired from the segmentation stage}
\end{flushleft} 
\begin{algorithmic}[2]
\scriptsize
\Procedure{PlotVesselWidth}{}
    \State Skeletonize the vessel mask
    \State Prune resultant centerline to remove unnecessary branches
    \For{every point on the centerline}
        \State Plot the normal to the centerline at that point
        \State Search for the two bounds lying on either side of the vessel mask on the normal line
        \State Calculate the euclidean distance between the bounds (this distance corresponds to a pixel width)
    \EndFor
    \State Plot graph of pixel widths against the point index at which they are measured
\EndProcedure
\end{algorithmic}
\end{algorithm}

\newpage
\section{Training details}
\label{appendix:training}
\begin{table}[!ht]
\centering
\scriptsize
\begin{threeparttable}
\caption{Training setup}
\centering
\begin{tabular}{ccc}
\hhline{===}
\renewcommand{\arraystretch}{1.6}
& Key frame extraction & Vessel segmentation\\
\hline
Batch size & 64 & 2\\
Weights initializer & random normal & HE normal\\
Optimizer & RAdam & Adam\\
LR & 1e-3 & 1e-4\\
$\beta_1$ & 0.9 & 0.9\\
$\beta_2$ & 0.999 & 0.999\\
Epsilon & 1e-8 & 1e-7\\

\hhline{===}
\end{tabular}
\end{threeparttable}
\end{table}

\begin{table}[!ht]
\centering
\scriptsize
\begin{threeparttable}
\renewcommand{\arraystretch}{1.3}
\caption{Augmentation hyperparameters for the key frame extraction model and the vessel segmentation model}
\label{table:augmentations}
\centering
\begin{tabular}{ccc}
\hhline{===}
Methods & Classification & Segmentation \\
\hline
Rotation angle & (\ang{-20},\ang{20}) & (\ang{-30},\ang{30}) \\
Shear angle & (\ang{-8},\ang{8}) & (\ang{-20},\ang{20}) \\
Horizontal translation & (0,0.2) & (-0.05,0.05) \\
Vertical translation & (-0.2,0.2) & (-0.3,0.3) \\
Scale & (0.8,1.0) & (0.7,1.3) \\
CLAHE & (1,5) & - \\
Sharpen & (0.4,0.6) & - \\
\hhline{===}
\end{tabular}
\begin{tablenotes}
\item [1]Values in the parentheses represent range. \item [2]Hyperparameters of translation and scale are fractions of the original size.
\end{tablenotes}
\end{threeparttable}
\end{table}
\end{document}